\newcommand{\diracslash}[1]{#1\llap{/\kern2pt}}

\newcommand{\be}{\begin{equation}}
\newcommand{\ee}{\end{equation}}
\newcommand{\bea}{\begin{eqnarray}}
\newcommand{\eea}{\end{eqnarray}}
\newcommand{\ba}[1]{\begin{array}{#1}}
\newcommand{\ea}{\end{array}}

\documentclass[prd,aps,floats,nofootinbib,floatfix,preprint]{revtex4}
\usepackage{graphicx}
\begin{document}

\title{Open charm and charmonium states in strong magnetic fields}

\author{Amruta Mishra}
\email{amruta@physics.iitd.ac.in}
\affiliation{Department of Physics, Indian Institute of Technology, Delhi,
Hauz Khas, New Delhi -- 110 016, India}

\author{S.P. Misra}
\email{misrasibaprasad@gmail.com}
\affiliation{Institute of Physics, Bhubaneswar -- 751005, India} 

\begin{abstract}
The mass modifications of the open charm ($D$ and $D^*$) mesons, 
and their effects on the decay widths $D^*\rightarrow D\pi$ 
as well as of the charmonium state, $\Psi(3770)$ to open charm
mesons ($\Psi(3770)\rightarrow D\bar D$), are investigated 
in the presence of strong magnetic fields. These are studied
accounting for the mixing of the pseudoscalar ($P$) and vector ($V$) 
mesons ($D-D^*$, $\eta_c'-\Psi(3770)$ mixings), with the mixing 
parameter, $g_{PV}$ of a phenomenological three-point 
($PV\gamma$) vertex interaction
determined from the observed radiative decay width of 
$V\rightarrow P\gamma$. 
For charged $D-D^*$ mixing, this parameter is dependent 
on the magnetic field, because of the Landau level 
contributions to the vacuum masses of these mesons.
The masses of the charged $D$ and $D^*$ mesons
modified  due to $PV$ mixing, in addition, have
contributions from the lowest Landau levels
in the presence of a strong magnetic field.
The effects of the magnetic field on the decay widths 
are studied using a field theoretic model of composite hadrons 
with quark (and antiquark) consittuents. 
The matrix elements for these decays are evaluated 
using the light quark--antiquark pair creation term 
of the free Dirac Hamiltonian for the constituent quark field,
with explicit constructions for the the charmonium state $\Psi(3770)$,
the open charm ($D$, $\bar D$, $D^*$) mesons
and the pion states in terms of the constituent quark fields.
The parameter for the charged $D-D^*$ mixing is observed 
to increase appreciably with increase in the magnetic field. 
This leads to dominant modifications to their masses, 
and hence the decay widths of charged $D^*\rightarrow D\pi$ as well as 
$\Psi(3770)\rightarrow D^+D^-$ at large values of the magnetic field.
The modifications of the masses and decay widths 
of the open and hidden charm mesons in the presence of strong magnetic
fields should have observable consequences on the production 
of the open charm ($D$ and $D^*$) mesons as well as 
of the charmonium states resulting from non-central 
ultrarelativistic heavy ion collision experiments.
\end{abstract}



\maketitle

\def\bfm#1{\mbox{\boldmath $#1$}}
\def\bfs#1{\mbox{\bf #1}}

\section{Introduction}
The study of heavy flavour hadrons  \cite{Hosaka_Prog_Part_Nucl_Phys} 
is a topic of extensive research in 
high energy physics. The topic has relevance in the high energy 
heavy ion collision experiments,
as the medium modifications of the properties of these hadrons 
can affect the experimental observables of heavy
ion collision experiments. There have been extensive 
studies of the heavy flavour mesons in the literature
using various frameworks. These include the potential models
\cite{eichten_1,eichten_2,Klumberg_Satz_Charmonium_prod_review,Quarkonia_QGP_Mocsy_IJMPA28_2013_review,repko},
the QCD sum rule approach
\cite{open_heavy_flavour_qsr,kimlee,klingl,amarvjpsi_qsr},
the coupled channel approach \cite{tolos_heavy_mesons},
a hadronic model using pion exchange \cite{sudoh},
a chiral effective model 
\cite{amarindam,amarvdmesonTprc,amarvepja,dpambmeson,amdpbottomonium}
and the quark meson coupling (QMC) model  
\cite{QMC_Krein_Thomas_Tsushima_Prog_Part_Nucl_Phys_2018}.
There are studies which show attractive interactions of the 
open heavy flavour ($\bar D$, $B$) mesons, as well as $J/\psi$ 
in nuclear matter, suggesting the interesting 
possibility of bound states of these mesons with nuclei
\cite{QMC_Krein_Thomas_Tsushima_Prog_Part_Nucl_Phys_2018}.
A study of the heavy quarkonium state
(with heavy quark and antiquark assumed to be bound by a Coulomb potential)
in the presence of a gluon field \cite{pes1,pes2,voloshin} 
shows that its mass shift is proportional to the change in the
gluon condensates in the medium. This is the result in the leading order
approximation, with the distance between the heavy quark and antiquark
assumed to be small as compared to the scale of gluonic fluctuations. 
The charmonium masses have been studied using the leading order
formula using a linear density approximation for the
gluon condensate \cite{leeko}. 

The estimation of strong magnetic fields produced 
in the peripheral ultra relativistic heavy ion
collision experiments \cite{Tuchin_Review_Adv_HEP_2013}
(of the order of $eB \sim 2 m_\pi^2$
at RHIC, BNL and $eB \sim 15m_\pi^2$ at LHC, CERN)
has initiated a lot of research activities in the study 
of the hadrons in the presence of magnetic fields.
The heavy flavour mesons are produced in the very 
early phase of the heavy ion collision experiments,
when the magnetic field can still be large. The effects
of the magnetic field can thus have consequences on the
observables of the heavy ion collision experiments.
The open heavy flavour mesons as well as the heavy quarkonium
states have been studied in the presence of a magnetic field
\cite{Gubler_D_mag_QSR,machado_1,B_mag_QSR,dmeson_mag,bmeson_mag,charmonium_mag,charmonium_mag_QSR,charmonium_mag_lee,Alford_Strickland_2013,Suzuki_Lee_2017}.
In the presence of a magnetic field, the mixing of the
pseudoscalar and the vector mesons has been shown to
lead to dominant contributions to the masses of these mesons 
within a QCD sum rule approach \cite{Gubler_D_mag_QSR,charmonium_mag_QSR,charmonium_mag_lee} as well as within an effective potential approach
\cite{Alford_Strickland_2013}. 

Using a chiral effective model, the mass modifications of the 
heavy flavour mesons in a hadronic medium have been studied
\cite{amarindam,amarvdmesonTprc,amarvepja,dpambmeson,amdpbottomonium}.
The masses of the open charm (bottom) mesons are modified 
due to their interactions with the baryons and scalar mesons
in the medium. On the other hand, the modifications of the charmonium
(bottomonium) are studied from the modification of 
the gluon condensates in the hadronic medium using the 
leading order QCD formula \cite{pes1,pes2,voloshin}.
This is calculated from the medium change of a dilaton field
which simulates within the effective hadronic model. 
Using a light quark pair creation model, namely
the $^3P_0$ model \cite{3p0,3p0_1}, as well as using 
a field theoretical model of composite hadrons 
with quark (and antiquark) constituents 
\cite{spm781,spm782,spmdiffscat}, 
the decay widths of the charmonium (bottomonium) to 
$D\bar D$ ($B\bar B$) in (magnetized) hadronic matter 
have also been studied from the mass modifications
of these mesons 
\cite{amarvepja,friman,amspmwg,amspm_upsilon,charm_decay_mag_3p0}.
In the presence of a magnetic field, the effects 
of the pseudoscalar meson - vector meson (PV)
mixings been studied for the open and hidden charm sector 
\cite{Gubler_D_mag_QSR,charmonium_mag_QSR,charmonium_mag_lee,Alford_Strickland_2013,charmonium_PV_amspm} and the strange meson sector
\cite{strange_PV_amspm}. 
These effects are found to have the dominant 
contributions to the masses of these mesons. 
$\Psi(3770)$ is the lowest charmonium state which can decay to
open charm mesons, $D\bar D$.
In the presence of strong magnetic fields, the decay width 
of $\Psi(3770)\rightarrow D\bar D$ arising from
the mass modifications of $\Psi(3770)$
due to PV ($\Psi(3770)-\eta_c'$) mixing and of the charged 
open charm mesons due to the contribution of Landau energy levels 
was studied in Ref. \cite{charmonium_PV_amspm}.
In the strange sector, the decay widths of $\phi\rightarrow K\bar K$
and $K^*\rightarrow K\pi$ 
in presence of strong magnetic fields \cite{strange_PV_amspm},
have been studied from the mass modificaitons of the $\phi$ and
the open strange mesons arising from PV mixing ($\phi-\eta '$ 
and $K-K^*$ mixings) and, in addition, the Landau level contributions 
for the charged $K$, $\bar K$ and $K^*$ mesons.
In the present work, the decay widths of
$D^*\rightarrow D\pi$ and $\Psi(3770)\rightarrow D\bar D$
are studied in the presence of a magnetic field
from the mass changes of the initial and final state mesons
accounting for the PV mixing ($\Psi(3770)-\eta_c'$, $D-D^*$) 
effects, as well as, accounting for the Landau level contributions
for the charged open charm mesons. 
These decay widths are computed using a field theoretic
model for composite hadrons with quark (and antiquark)
constitutents.

The outline of the paper is as follows. In section II,
the effects of the magnetic field on the masses of the
$D$ and $D^*$ mesons are studied. 
In  the presence of a magnetic field, the
mass modifications of these mesons arise due the
mixing of the pseudoscalar, $D$ mesons with the 
longitudinal component of the  vector, $D^*$ meson.
The charged $D$ and $D^*$ mesons have additional
contributions from the Landau quantization in the
presence of the magnetic field. In section III, the effects of the
magnetic field on the  $D^* \rightarrow D\pi$
as well as $\Psi(3770)\rightarrow D\bar D$ 
decay widths
are investigated using a field theoretic model of
composite hadrons with quark (and antiquark) constituents. 
These decay widths are calculated
using the light quark antiquark pair creation term of the
free Dirac Hamiltonian and explicit constructions for these
mesons. In section IV, we discuss 
the results of the effects of the magnetic field
on the masses of the open charm ($D$ and $D^*$) mesons, 
and their subsequent effects on the decay widths 
$D^*\rightarrow D\pi$ and $\Psi(3770)\rightarrow D\bar D$.
For the charmonium decay width, 
contributions to the masses of the $\Psi(3770)$ as
well as open charm mesons due to  
$D-D^*$ and $\Psi(3770)-\eta_c'$ mixings
are taken into account, and, in addition, the Landau level
contributions for the charged open charm mesons. 
In section V, we summarize the findings of the present study.

\section{Masses of $D$ and $D^*$ mesons in presence of a magnetic field}

In this section, we investigate the effects of a uniform 
magnetic field on the masses of the open charm
($D$ and $D^*$) mesons. The modifications in the masses
of these mesons arise from the mixing of the pseudoscalar ($D$)
and the vector ($D^*$) mesons in the presence
of an external magnetic field, with additional contributions
from the Landau energy levels for the charged open
charm mesons. 

The mixing of the pseudoscalar ($D$) and vector ($D^*$) mesons
in the presence of a magnetic field 
is taken into account through the interaction 
\cite{charmonium_mag_lee,Gubler_D_mag_QSR} 
\begin{equation}
{\cal L}_{PV\gamma}=\frac{g_{PV}}{m_{av}} e {\tilde F}_{\mu \nu}
(\partial ^\mu P) V^\nu,
\label{PVgamma}
\end{equation}
where $m_{av}=(m_V+m_P)/2$, $m_P$ and $m_V$ are the masses 
for the pseudoscalar and vector charmonium states.
In equation (\ref{PVgamma}), the coupling parameter $g_{PV}$
is fitted from the observed radiative decay width 
$\Gamma(V\rightarrow P +\gamma)$, given as
\begin{equation}
\Gamma (V\rightarrow P \gamma)
=\frac{e^2}{12}\frac{g_{PV}^2 {p_{cm}}^3}{\pi m_{av}^2}.
\label{decay_VP}
\end{equation}
In the above, $p_{cm}$ is the magnitude of the center of mass
momentum in the final state given as
$p_{cm}=(m_V^2-m_P^2)/(2m_V)$.
The modified masses due to the mixing of the pseudoscalar 
and the longitudinal component of the vector mesons, 
are given by
\begin{equation}
m^{(PV)}_{V^{||},P}=\frac{1}{2} \Bigg ( M_+^2 
+\frac{c_{PV}^2}{m_{av}^2} \pm 
\sqrt {M_-^4+\frac{2c_{PV}^2 M_+^2}{m_{av}^2} 
+\frac{c_{PV}^4}{m_{av}^4}} \Bigg),
\label{mpv_long}
\end{equation}
where $M_+^2=m_P^2+m_V^2$, $M_-^2=m_V^2-m_P^2$ and 
$c_{PV}= g_{PV} eB$.

In the presence of an external magnetic field,
the masses of the charged $D$ and $D^*$ mesons 
have contributions from the Landau levels.
In the present work, the mass of the charged 
open charm meson is taken to be the
energy of the ground state (lowest Landau level) 
at $p_z=0$, which is a valid approximation 
for strong magnetic fields. However,
for weak magnetic fields,
the Landau energy levels of the excited states
of the charged mesons are quite close to the
the ground state and the mass of the charged meson
has contributions also from the excited Landau levels.
A consistent summation of all the Landau levels
has been carried out in the study of the charged
$D$ mesons in the presence of a magnetic field
using the QCD sum rule approach in Ref. \cite{Gubler_D_mag_QSR}.
In the present study, the masses of the charged 
open charm mesons are taken to be arising from
the lowest Landau level ($n=0$) as
\begin{equation}
m^{eff}_{D^\pm}=\sqrt {{m_{D^\pm}}^2 +eB},\,\,\;\;\;\;
m^{eff}_{{D^*}^\pm}=\sqrt {{m_{{D^*}^\pm}}^2 -eB},
\label{m_LL}
\end{equation}
where the value of the gyromagnetic ratio of $D^*$ meson 
has been taken to be 2.
The mixing parameter for charged $D-D^*$ mixing,
determined from the observed decay width $D^*\rightarrow D\gamma$,
is magnetic field dependent, because of the lowest Landau  
level contributions to their vacuum masses.
The masses of these charged mesons modified 
due to the mixing effect in the presence
of a magnetic field, are obtained from
equation (\ref{mpv_long}), using the magnetic field
dependent $PV$ mixing parameter, and 
the values of the masses of the charged pseudoscalar 
and vector mesons ($m_P$ and $m_V$) taken as the
lowest Landau level contribution to their vacuum masses.
The masses of the charged mesons modified due to 
$PV$ mixing, in addition, have contributions from the lowest
Landau levels. 

The in-medium decay width of $D^*\rightarrow D\pi$
arising from the mass modifications of the $D$ and $D^*$ mesons,
and of $\Psi(3770)\rightarrow D\bar D$, due to mass
modifications of the $\Psi(3770)$ as well as $D$ and $\bar D$
mesons, are investigated in the presence of strong magnetic fields. 
The $PV$ mixing is observed to lead to 
a drop (rise) in the masses of the $D$ (longitudinal $D^*$)
mesons, and it is observed to be quite 
appreciable for the neutral open charm mesons
\cite{Gubler_D_mag_QSR}, as compared to the case
of charged open charm mesons, when the $PV$ mixing 
parameter, $g_{PV}$ is kept at the zero magnetic field
value. However, as we shall see later, the strong increase 
of this parameter with the magnetic field, is observed 
to lead to significant modifications to the masses
of the charged open charm mesons, and hence to the decay
widths of the charged $D^*\rightarrow D\pi$ as well as
$\Psi(3770)\rightarrow D^+D^-$.   
There is substantial modification in the mass of the longitudinal component
of $\Psi(3770)$ is due to mixing with $\eta_c'$ in the 
presence of a magnetic field, which was observed to lead
to appreciable modification to the decay width of 
$\Psi(3770)\rightarrow D\bar D$ \cite{charmonium_PV_amspm}.
Due to the Landau level contributions for the charged open charm mesons 
the decay widths for the $D^+D^-$ and $D^0 {\bar {D^0}}$ final states
are observed to be quite different in the presence of a magnetic field
\cite{charmonium_PV_amspm}. 
In the present work, the decay widths 
of $\Psi(3770)\rightarrow D\bar D (D^+D^-,D^0{\bar {D^0}})$
are studied accounting for the PV mixing contributions to the masses
of the $D$ and $\bar D$ mesons (arising
from the $D-D^*$ and $\bar D-{\bar D^*}$ mixings), and, 
in addition the Landau level contributions for the charged 
open charm mesons. The decay widths for $D^*\rightarrow D\pi$ and
$\Psi(3770)\rightarrow D\bar D$ are calculated using a field theoretical
model of composite hadrons with quark (and antiquark) 
constituents as described in the following section.

\section{Two body Decay widths within a model for composite hadrons}
\label{dwFT_comp_had}

The decay width for a generic two body decay process 
of $A ({\bf 0})\rightarrow B({\bf p}) + C(-{\bf p})$ 
is calculated 
within the field theoretic model of composite hadrons 
with explicit contructions for the
initial and final states in terms of the quark (and antiquark)
constituents and using the matrix element of free Dirac Hamiltonian 
as the  light quark pair creation, between the initial and final
states. In the present work, we calculate the decay widths
for $D^*\rightarrow D\pi$ and $\Psi(3770)\rightarrow D\bar D$
accounting for the PV mixing contributions for the masses of
the open charm mesons and charmonium states 
in the presence of a magnetic field. 

\subsection{Decay width of $D^*\rightarrow D\pi$}

The effects of a uniform background magnetic field on the decay widths
of ${D^*}\rightarrow D\pi$ 
are studied from the mass modifications of the
$D$ and $D^*$ mesons in the presence of the magnetic field. 
The decay widths of the charged ${D^*}^+$ meson 
(${D^*}^+\rightarrow D^+\pi^0$ and ${D^*}^+\rightarrow D^0\pi^+$), 
and the neutral ${D^*}^0$ meson (${D^*}^0\rightarrow D^0\pi^0$),  
are computed using a field theoretical model with composite 
hadrons \cite{amspmwg}, comprising of quark (and antiquark) 
constituents \cite{spm781,spm782,spmdiffscat}.
Using a Lorentz boosting, the constituent quark field operators 
of the hadron in motion are obtained from the constituent quark 
field operators of the hadron at rest.
In the present model for the composite hadrons it is 
assumed that the quark (and antiquark) constituents carry 
fractions of the energy of the hadron 
\cite{spm781,spm782}.
This is similar to the MIT bag model \cite{MIT_bag}, 
where the quarks (antiquarks) occupy
specific energy levels inside the hadron. 

The decay width for $D^*\rightarrow D\pi$
is obtained from the matrix element of the quark-antiquark
pair creation term of the free Dirac Hamiltonian in terms 
of the constituent quark field operators between the
initial and final states. 
Assuming harmonic oscllator wave functions for these states,
the explicit constructions for the decaying $D^*$ meson at rest 
and the produced states, $D$ and $\pi$ with finite momenta,
are given as

\begin{eqnarray}
|{D^*}^m ({\bf 0})\rangle & =& 
\frac{1}{\sqrt{6}}
\Bigg (\frac {R_{D^*}^2}{\pi} \Bigg)^{3/4}
\int d{\bf k} 
\exp\Bigg(-\frac {R_{D^*}^2 {\bf k}^2}{2}\Bigg)
{c_r}^{i}({\bf k})
^\dagger u_r^\dagger \sigma^m 
\tilde {q_s}^{i} 
(-{\bf k})v_s
d\bfs k |vac\rangle,
\label{dstr}
\\
|D ({\bf p})\rangle & = &
\frac{1}{\sqrt{6}}
\Bigg (\frac {R_D^2}{\pi} \Bigg)^{3/4}
\int d{\bf k} 
\exp\Bigg(-\frac {R_D^2 {\bf k}^2}{2}\Bigg)
{c_r}^{i}({\bf k}+\lambda_2 {\bf p})
^\dagger u_r^\dagger 
\tilde {q_s}^{i} 
(-{\bf k} +\lambda_1 {\bf p})v_s
d\bfs k |vac\rangle,
\label{d}
\\
|\pi^+ ({\bf p})\rangle & =& 
\frac{1}{\sqrt{6}}
\Bigg (\frac {R_\pi^2}{\pi} \Bigg)^{3/4}
\int d{\bf k} 
\exp\Bigg(-\frac {R_\pi^2 {\bf k}^2}{2}\Bigg)
{u_r}^{i}({\bf k} +\frac {\bf p}{2})
^\dagger u_r^\dagger 
\tilde {d_s}^{i} 
(-{\bf k} +\frac {\bf p}{2})
v_s
d\bfs k |vac\rangle,
\label{pip}
\\
|\pi^0(\bfs p)> &=&
\frac{1}{2\sqrt 3}
\left(\frac{R_{\pi}^2}{\pi}\right)^{3/4}
\int d\bfs k 
\exp\Bigg(-\frac{1}{2}R_\pi^2{\bfs k}^2\Bigg)
\Bigg ({u_r}^{i}({\bf k} +\frac {\bf p}{2})
^\dagger u_r^\dagger 
\tilde {u_s}^{i} 
(-{\bf k} +\frac {\bf p}{2}) v_s
\nonumber \\
&-&{d_r}^{i}({\bf k} +\frac {\bf p}{2})
^\dagger u_r^\dagger 
\tilde {d_s}^{i} 
(-{\bf k} +\frac {\bf p}{2}) v_s \Bigg)
|vac>,
\label{pi0}
\end{eqnarray}
In the above equations,
${c_r^i}({\bfs k})^\dagger ( {{\tilde c}_r^i}({\bfs k}))$
is the creation operator of the heavy flavour charm quark 
(antiquark) with spin $r$, color $i$ and momentum ${\bf k}$,
${q_r^i}({\bfs k})^\dagger ( {{\tilde q}_r^i}({\bfs k}))$
refer to the light ($q=(u,d)$) quark (antiquark),
and $u_r$ and $v_s$ are the two component spinors.
In equations (\ref{dstr}) and (\ref{d}), $q=d (u)$ correspond
to ${D^*}^+ ({D^*}^0)$ and  ${D}^+ ({D}^0)$ states respectively. 
The parameters $R_{D^*}$, $R_D$ and $R_\pi$ refer  to 
the harmonic oscillator strengths for the states $D^*$,
$D$ and $\pi$ respectively.
In equation (\ref{d}), $\lambda_1$ and $\lambda_2$ are the 
fractions of the mass (energy) of the $D$ meson
at rest (in motion), carried by the constituent 
light (d,u) antiquark and the constituent heavy charm quark,
with $\lambda_1 +\lambda_2=1$. These are calculated 
by assuming the binding energy of the hadron 
as shared by the quark (antquark) to be inversely 
proportional to the quark (antiquark) mass
\cite{spm782}. For the pion states, $\pi^+$ and $\pi^0$, 
which are light quark-antiquark bound states, the fractions
of energy carried by the quark and antiquark are the same,
i.e., $\lambda_1=\lambda_2=1/2$, as the masses of the $u$
and $d$ quarks are assumed to be the same.

The decay width for the process $D^*({\bf 0}) \rightarrow 
D({\bf p})+\pi ({\bf p'})$, is obtained from
the matrix element of the light quark pair creation term of
the free Dirac Hamiltonian density
\begin{equation}
{\cal H}_{q^\dagger\tilde q}({\bf x},t=0)
=Q_{q}^{(p)}({\bf x})^\dagger 
(-i
\mbox{\boldmath $\alpha$}\cdot
\mbox{\boldmath $\bigtriangledown$} 
+\beta M_q)
{\tilde Q}_q^{(p')}({\bf x}) 
\label{hint}
\end{equation}
between the initial and the final states.
In equation (\ref{hint}),
$M_q$ is the constituent mass of the light quark, 
$q=(u,d)$, $\mbox {\boldmath $\alpha$}$ and $\beta$ are the Dirac
matrices, and, the subscript $q$ of the field operators 
in equation (\ref{hint})  
refers to the fact that the $\bar q$ and $q$ are the constituents 
of the $D$ and $\pi$ mesons with momenta ${\bf p}$ and
${\bf p'}$ respectively in the final state of the
decay of the $D^*$ meson.
The matrix element is evaluated using the
explicit contructions for the inital and final states 
given by equations (\ref{dstr})--(\ref{pi0}) and is 
obtained as   
\begin{equation}
\langle D ({\bf p}) | \langle \pi ({\bf p}')|
\int{\cal H}_{q^\dagger\tilde q}({\bf x},t=0)d{\bf x}
|{D^*}^m ({\bf 0}) \rangle
=\delta({\bf p}+{\bf p}') A_{D^{*}}({|\bf p|}) {\bf p}^m,
\label{dstardpi}
\end{equation}
where,
\begin{eqnarray}
A^{D^*}(|{\bf p}|)&=&6c
\Bigg(\frac{\pi}{a}\Bigg)^{{3}/{2}}
\exp\left[ab^2{|\bf p|}^2
-\frac{1}{2}\left(\lambda_2^2 R_D^2
+\frac{1}{4}R_\pi^2\right){|\bf p|}^2\right] 
\Big[F_0+F_1 \Big (\frac{3}{2a}\Big )\Big].
\label{apdstr}
\end{eqnarray}
In the above, 
\begin{eqnarray}
&&a=\frac{1}{2}\left(R_{D^*}^2+R_D^2+R_\pi^2\right); \;\;\;\; 
b=\frac{1}{2a}\left(R_D^2\lambda_2
+\frac{1}{2}R_\pi^2\right),\;\;\;\;
c=\frac{1}{6}\cdot \frac{1}{2\sqrt 3}
\left(\frac{R_{D^*}^2 R_D^2 R_\pi^2}{\pi^3}\right)^{\frac{3}{4}},
\label{abcdstr}
\end{eqnarray}
and,
\begin{eqnarray}
F_0&=&(b-1)\left(1-\frac{1}{8M_q^2}{|\bfs p|}^2
(\lambda_2-\frac{1}{2})^2\right) \nonumber \\
&-&(b-\lambda_2)\left(\frac{1}{2}+\frac{1}{4M_q^2}{|\bfs p|}^2
\left(\frac{3}{4}b^2-\frac{5}{4}b
+\frac{7}{16}\right)\right)
\nonumber\\
&-&(b-\frac{1}{2})\left[\frac{1}{2}+\frac{1}{4M_q^2}{|\bfs p|}^2
\left(\frac{3}{4}b^2-(1+\frac{1}{2}\lambda_2)b
+\lambda_2-\frac{1}{4}\lambda_2^2\right)\right]\nonumber\\
F_1&=&-\frac{1}{4M_q^2}\left[\frac{5}{2}b-\frac{9}{8}
-\frac{11}{12}\lambda_2\right].
\label{c01dstr}
\end{eqnarray}
For the $D^*$ meson decaying at rest, the magnitude of the
3-momentum of the outgoing $D (\pi$) meson is given as
\begin{equation}
|\bfs p|=\Bigg (\frac{1}{4}m_{D^{*}}^2-\frac{m_{D}^2+m_{\pi}^2}{2}
+\frac{\left(m_{D}^2-m_{\pi}^2\right)^2}{4m_{D^{*}}^2}
\Bigg )^{1/2}.
\label{modp}
\end{equation}
With $\langle f | S |i\rangle =\delta_4 (P_f-P_i) M_{fi}$,
we have
\begin{equation}
M_{fi}=2\pi (-i A^ {D^*}  (|{\bf p}|)p^m.
\end{equation}
The expression of the decay width of 
$D^* \rightarrow D\pi$ is obtained 
by taking the average 
over the initial spin components as
\begin{eqnarray}
\Gamma\left(D^{*} ({\bf 0})\rightarrow 
D ({\bf p}+\pi ({-\bf p})\right)
&&= \gamma_{D^*}^2\frac{1}{2\pi}
\int\delta(m_{D^{*}}-p_D^0-p_\pi^0)|M_{fi}|_{av}^2d\bfs p
\nonumber\\
&=& \gamma_{D^*}^2\frac{8\pi^2p_D^0p_\pi^0}{3m_{D^*}}|A^{D^*}
(|\bfs p|)|^2|\bfs p|^3,
\label{gammadstr}
\end{eqnarray}
where $p_D^0=(|{\bf p}|^2+m_D^2)^{1/2}$ and 
$p_\pi^0=(|{\bf p}|^2+m_\pi^2)^{1/2}$ are the energies of
the outgoing $D$ meson and pion respectively.
In the above equation, $\gamma_{D^*}$ is the production strength
of $D\pi$ from decay of $D^*$ meson, which is fitted from
its vaccum decay width. In the presence of the magnetic field,
the masses of the charged $D$ and $D^*$ mesons 
the lowest Landau level are given by equation (\ref{m_LL}).
The expression for the
decay width of $D^*\rightarrow D\pi$ given by equation 
(\ref{gammadstr}) does not account for
the mixing of the $D$ and $D^*$ mesons in the
presence of magnetic fields. The mixing of the pseudoscalar
mesons and the vector mesons leads to a drop (increase)
in the mass of the $D$ meson (longitudinal component 
of the $D^*$ meson) in the presence of a uniform magnetic
field. 
When we include the mixing effect, the expression
for the decay width of $D^*\rightarrow D\pi$ is modified to
\begin{eqnarray}
&&\Gamma^{PV}(D^* ({\bf 0}) \rightarrow  
D({\bf p}) {\pi} (-{\bf p}))
=\gamma_{D^*}^2\frac{8\pi^2}{3}
\Bigg [ 
\Bigg(\frac{2}{3} |{\bf p}|^3 
\frac {p^0_D (|{\bf p}|) p^0_{\pi}(|{\bf p}|)}{m_{D^*}}
A^{D^*}(|{\bf p}|)^2 \Bigg)
\nonumber \\
&+&\Bigg(\frac{1}{3} |{\bf p}|^3 
\frac {p^0_D(|{\bf p}|) p^0_{\pi}(|{\bf p}|)}{m_{D^*}^{PV}}
A^{D^*}(|{\bf p}|)^2 \Bigg) \Big({|{\bf p}|\rightarrow |{\bf p|}
(m_{D^*} = m_{D^*}^{PV}, m_{D} = m_{D}^{PV}
)}\Big)
\Bigg]. 
\label{gammadstr_mix}
\end{eqnarray}
In the above, the first term corresponds to the transverse
polarizations for the vector $D^*$ meson, which 
remain unaffected by the mixing of the $D$ and $D^*$
states. However, the masses of the charged $D^*$
(and $D$) mesons are modified due to the lowest Landau
level as given by equation (\ref{m_LL}). 
The second term in (\ref{gammadstr_mix})
corresponds to the longitudinal component of the
$D^*$ meson, where the pseudoscalar-vector meson 
mixing leads to the modifications of the masses
of the longitudinal component of the $D^*$ meson
as well as the $D$ mesons,
as given by equation (\ref{mpv_long}).
In the presence of the magnetic field,
these mixing effects are taken into account
in addition to the Landau level contributions
for the charged $D$ and $D^*$ mesons in the 
presence of the magnetic fields. For the charged
mesons the transverse components are also 
modified in the presence of the magnetic field,
whereas, the mixing effects modify only the masses
of the longitudinal component of the $D^*$ meson.

\subsection{Decay width of $\Psi(3770)\rightarrow D\bar D$}

The decay width of
$\Psi(3770)\rightarrow D\bar D$ in (magnetized) matter 
was investigated using the field theoretical model
for hadrons with quark (and antiquark) constituents
\cite{amspmwg,charmonium_PV_amspm}.
The expression of the decay width of $\Psi(3770)\rightarrow D\bar D$ 
including the PV ($\Psi(3770)-\eta_c'$) mixing in the presence 
of a magnetic field is given as \cite{charmonium_PV_amspm}
\begin{eqnarray}
\Gamma^{PV}(\Psi &\rightarrow & D({\bf P}) {\bar D} (-{\bf P}))
=\gamma_\Psi^2\frac{8\pi^2}{3}
\Bigg [ 
\Bigg(\frac{2}{3} |{\bf P}|^3 
\frac {P^0_D (|{\bf P}|) P^0_{\bar D}(|{\bf P}|)}{m_\Psi}
A^{\Psi}(|{\bf P}|)^2 \Bigg)
\nonumber \\
&+&\Bigg(\frac{1}{3} |{\bf P}|^3 
\frac {P^0_D(|{\bf P}|) P^0_{\bar D}(|{\bf P}|)}{m_{\Psi}^{PV}}
A^{\Psi}(|{\bf P}|)^2 \Bigg) \Big({|{\bf P}|\rightarrow |{\bf P|}
(m_\Psi = m_\Psi^{PV})}\Big)
\Bigg], 
\label{gammapsiddbar_mix}
\end{eqnarray}
where,
\begin{eqnarray}
A^{\Psi}(|{\bf P}|) = 6c_\Psi\exp[(a_\Psi {b_\Psi}^2
-R_D^2\lambda_2^2){\bf P}^2]
\cdot\Big(\frac{\pi}{a_\Psi}\Big)^{{3}/{2}}
\Big[F_0^\Psi+F_1^\Psi\frac{3}{2a_\Psi}
+F_2^\Psi\frac{15}{4a_\Psi^2}\Big].
\label{ap}
\end{eqnarray}
The parameters $a_\Psi$, $b_\Psi$ and $c_\Psi$  
are given in terms of $R_D$ and $R_\Psi$, which are
the strengths of the harmonic oscillator wave functions
for the $D(\bar D)$ and the charmonium states,
and $F_i^\Psi (i=0,1,2)$ are polynomials in $|{\bf P}|$, 
the magnitude of the momentum of the outgoing $D(\bar D)$ meson
given by \cite{amspmwg} 
\begin{equation}
|{\bf P}|=\Big (\frac{{m_\Psi}^2}{4}-\frac {{m_D}^2+{m_{\bar D}}^2}{2}
+\frac {({m_D}^2-{m_{\bar D}}^2)^2}{4 {m_\Psi}^2}\Big)^{1/2}.
\label{ppd}
\end{equation}
The first term in equation (\ref{gammapsiddbar_mix}) 
corresponds to the transverse
polarizations for the charmonium state, $\Psi\equiv \Psi(3770)$, 
whose masses remain unaffected by the mixing of the 
pseudoscalar and vector charmonium
states. The second term in (\ref{gammapsiddbar_mix})
corresponds to the longitudinal component of the
charmonium state whose mass is modified due to mixing
with the pseudoscalar meson in the presence of the 
magnetic field \cite{charmonium_PV_amspm}.
In Ref. \cite{charmonium_PV_amspm}, the mass modification 
of $\Psi(3770)$ due to mixing with
$\eta_c'$ and the Landau level contributions for $D$ and
$\bar D$ were considered for the computation of the
the in-medium decay width in the presence of a magnetic field.
In the present work,
we compute the decay width accounting also for mass modifications
of $D$ and $\bar D$ due to mixings with the vector $D^*$
and $\bar {D^*}$ mesons in the presence of a magnetic field.

\section{Results and Discussions}

The masses of the $D$ and $D^*$ are calculated in the 
presence of a uniform magnetic field arising due to the
mixing of the pseudoscalar and vector mesons, with 
additional Landau level contributions 
for the charged open charm mesons. The mixing is taken
into account through a phenomenological Lagrangian
interaction given by equation (\ref{PVgamma}). This is
observed to lead to substantial drop (rise) in the
pseusoscalar (longitudinal component of the vector) meson.
The coupling parameter $g_{PV}$ is calculated from
the radiative decay width of the vector meson, $V$
to pseudoscalar meson, $P$, $V\rightarrow P\gamma$.
There is observed to be mass splitting of the $D^0$ 
and ${D^*}^0$, as well as, for $D^+$ and ${D^*}^+$
due to the mixing. As we shall see later, the effect 
is observed to be much more prominent for the neutral 
mesons as compared to the charged open charm mesons,
when the $g_{PV}$ is fixed
at its the zero magnetic field value. 
This is due to the much larger value of the
mixing strength parameter, $g_{PV}$ for the neutral mesons
(${D^*}^0$ and ${D}^0$) as compared to $g_{PV}(eB=0)$
for the charged mesons (${D^*}^+$ and ${D}^+$), 
However, as has been mentioned earlier, the
mixing parameter for the charged mesons is magnetic
field dependent, because of the Landau level contributions
to the vacuum masses of these mesons. The value
of the mixing parameter for the charged open charm mesons
is observed to increase appreciably at high 
magnetic fields, which leads to significant modifications
to the masses of these mesons for large magnetic fields.

\begin{figure}
\includegraphics[width=15cm,height=15cm]{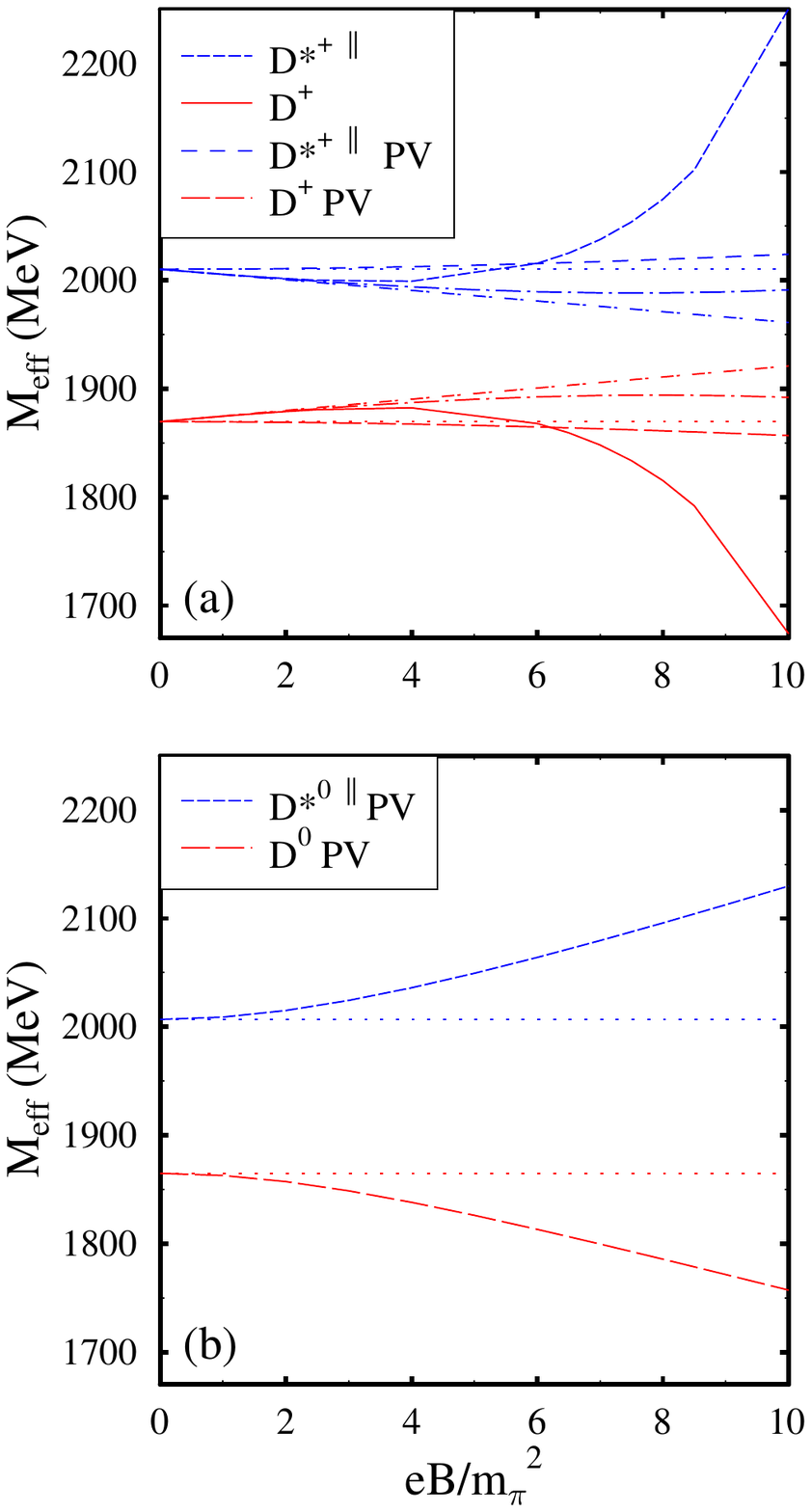}
\caption{(Color online)
The masses of the $D$ meson and the longitudinal component 
of the $D^*$ mesons are plotted as functions of $eB/{m_\pi^2}$. 
The masses of the charged mesons
${D^*}^+$ and $D^+$ are compared with the masses for the 
case when the PV mixing parameter is fixed at the zero magnetic field
value (shown as long dash-dotted lines). 
The effects of the pseudoscalar--vector (PV) mixing on these
masses are shown for the charged (for $g_{PV}=g_{PV}(eB=0)$)
and neutral mesons. The Landau contributions to the masses
of the charged mesons are also shown as the short dash-dotted
lines. The dotted lines show the masses when the mixing effects
as well as the Landau level contributions (for charged mesons)
are not included. 
}
\label{md_PV}
\end{figure}

\begin{figure}
\includegraphics[width=15cm,height=15cm]{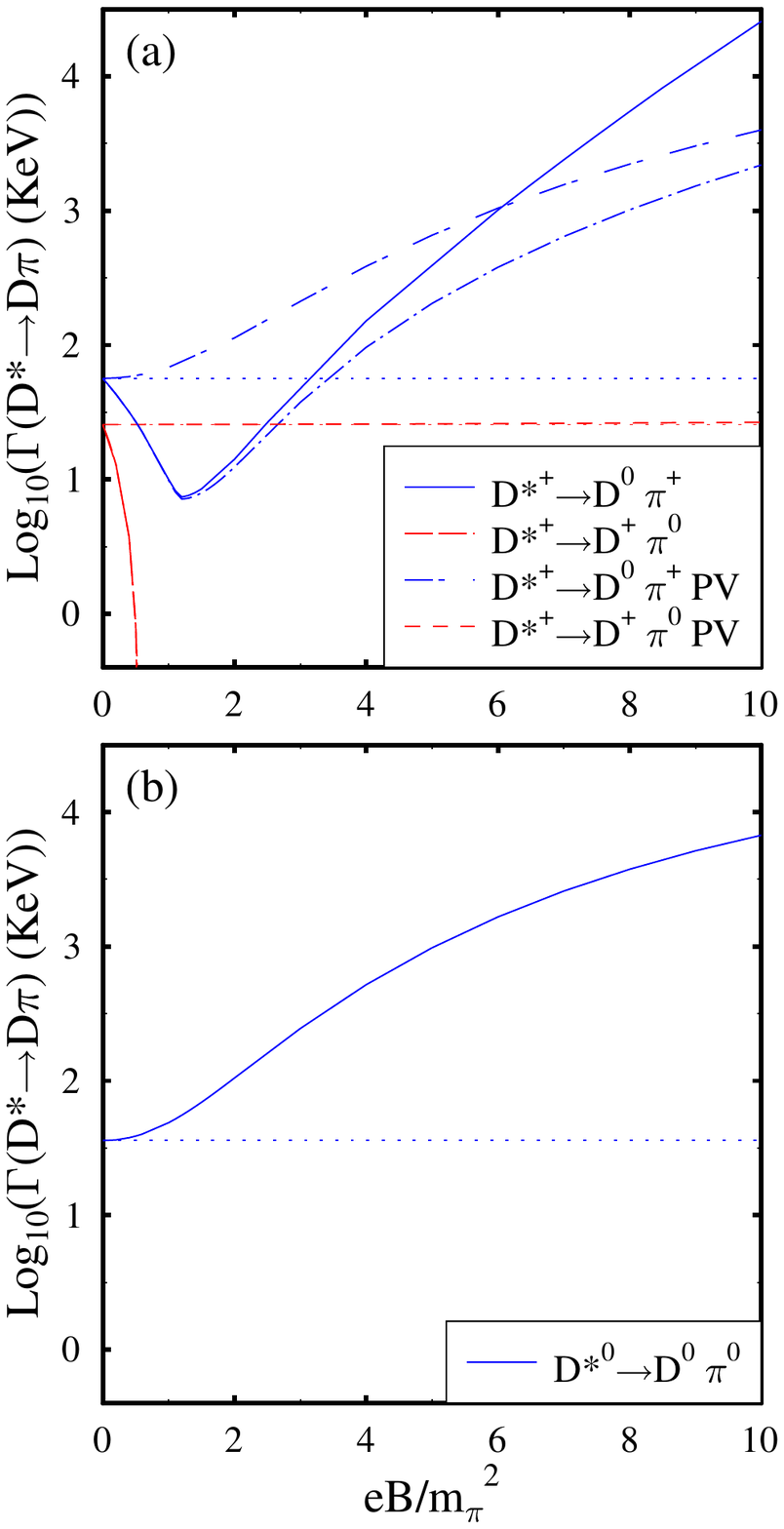}
\caption{(Color online)
The logarithm of the decay widths for $D^*\rightarrow D\pi$
(in KeV) for the charged ${D^*}^+$ and neutral ${D^*}^0$ are 
plotted as functions of $eB/{m_\pi^2}$
in panels (a) and (b) respectively. The PV mixing parameter
of the charged mesons is magnetic field dependent due to
Landau level contributions to their masses. 
The decay widths, ${D^*}^+\rightarrow D^0\pi^+$
as well as ${D^*}^+\rightarrow D^+\pi^0$ are shown accounting for
the magnetic field dependence of this parameter, and compared with
the cases when this parameter is fixed at the zero magnetic field 
value (shown as the short dash-dotted lines). The decay widths 
are also plotted for the case with only the PV mixing effects 
(with $g_{PV}=g_{PV}(eB=0)$).
}
\label{dwFT_Dstr}
\end{figure}

\begin{figure}
\includegraphics[width=15cm,height=15cm]{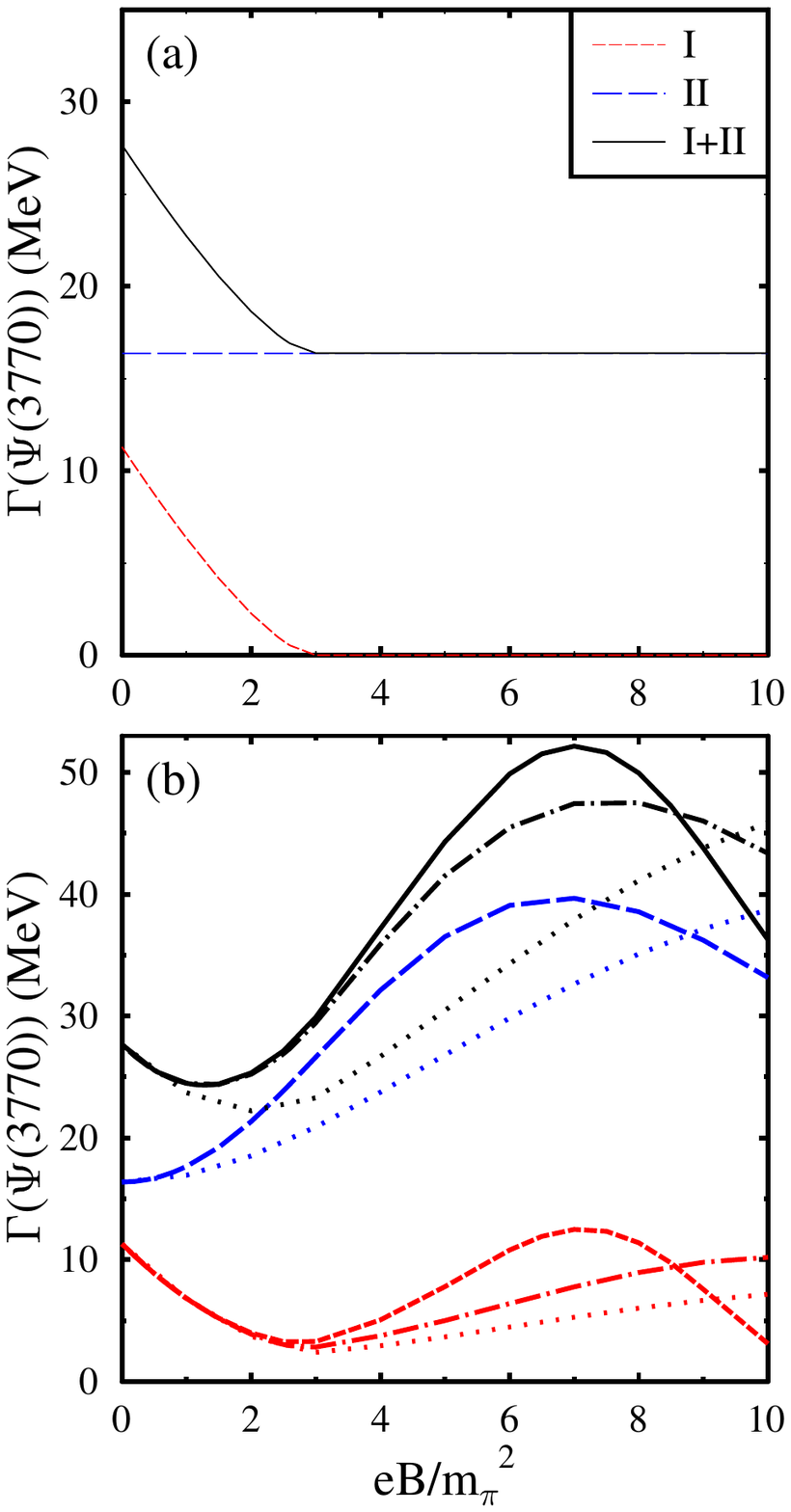}
\caption{(Color online)
The partial decay widths for $\Psi(3770)\rightarrow D\bar D$
(in MeV) plotted as functions of $eB/{m_\pi^2}$
for the final states (I) $D^+D^-$, (II) $D^0 {\bar {D^0}}$
and total of these two contributions (I+II). These are
plotted without and with the pseudoscalar-vector meson (PV) mixing 
effects on the masses of the initial and final state
mesons in panels (a) and (b) respectively.
In panel (b), the results are compared with the
case (shown as dotted lines) when the mass modifications 
of the $D$ and $\bar D$ mesons do not include the 
contributions from the PV mixing.
The decay width for (I) $\Psi(3770)\rightarrow D^+D^-$
as well as the total (I+II) are compared to the case when the 
${D^\pm}-{D^*}^\pm$ mixing parameter is fixed at the zero 
magnetic field value (shown as dot-dashed lines).
}
\label{dwFT_3770_vac_ddst_spinmix}
\end{figure}

In the present work, we investigate the effects 
of the magnetic field on the decay widths
$D^*\rightarrow D\pi$ from the mass modifications
of the $D^*$ and $D$ mesons in the presence
of a magnetic field. The decay widths for the charged
$D^*$ meson (${D^*}^+ \rightarrow D^+ \pi^0$ and
${D^*}^+ \rightarrow D^0 \pi^+$) and the neutral
${D^*}^0$ meson (${D^*}^0 \rightarrow D^0 \pi^0$) 
are calculated using a field theoretical model
for composite hadrons as described in the previous
section. The values of the constituent light quark
($q=u,d$) masses are taken to be $M_u=M_d=330$ MeV
and that of the heavy charm quark as 1600 MeV \cite{amspmwg}. 
The vacuum values for the masses (in MeV) of these mesons
are taken to be  $m_{{D^*}^+}=2010.26,\,\,m_{{D^*}^0}=2006.85,\,\,
m_{{D}^+}=1869.65,\,\,m_{{D}^0}=1864.8,\,\,m_{{\pi}^+}=139.57,\,\,
m_{{\pi}^0}=135$ \cite{pdg_2019_update}.
As has already been mentioned, the parameters
$\lambda_1$ and $\lambda_2$ in equation (\ref{d}), 
are the fractions of the mass (energy) of the $D$ meson
at rest (in motion), carried by the constituent 
light antiquark and the constituent heavy charm quark.
For the quark-antiqaurk bound states, these fractions
add up to 1, i.e., $\lambda_1+\lambda_2=1$,
yielding the sum of the masses (energies) of the
quark and antiquark constituents to be the mass (energy)
of the heavy flavour meson. 
The value of $\lambda_2$ is calculated to be 0.85 
\cite{amspmwg},
by assuming the binding energy of the hadron 
as shared by the quark (antquark) to be inversely 
proportional to the quark (antiquark) mass
\cite{spm782}. The harmonic oscillator strength  
parameter for $D$ meson is taken as $R_D$=(310 MeV)$^{-1}$ 
which is consistent with the experimental
values of the vacuum decay widths of 
$\Psi (3770)\rightarrow D\bar D$,
and, $\Psi(4040)$ to $D\bar D$, $D\bar D^*$, $D^* \bar D$ and 
$D^*\bar D^*$ \cite{leeko,amspmwg}. 
In the present work of computation of the decay width,
$\Gamma (D^* \rightarrow D\pi)$, the value of the harmonic oscillator 
strength for $D^*$ meson, $R_{D^*}$ has been taken to be the same as
that of the $D$ meson, $R_D$ and the value of $R_{\pi}$ 
of the pion wave function is fitted to the square of the
charge radius of pion as $0.4 {\rm fm}^2$, which yields
$R_\pi$=(211 MeV)$^{-1}$ \cite{spm782,amspmwg}. 

The mixing of the $D$ and $D^*$ in the presence of a
magnetic field is described by the interaction Lagrangian
given by equation (\ref{PVgamma}). The coupling 
constant, $g_{PV}$ is obtained from the observed decay width
for the process $V\rightarrow P \gamma$. For the process
$D^{*+} \rightarrow D^+ \gamma$, the observed value of 
1.33 keV (1.6\% of the total width of ${D^*}^+$ 
of 83.4 KeV) (which is the same as the decay width
of $D^{*-} \rightarrow D^- \gamma$)
gives the value of the coupling constant,
$g_{D^{\pm}{D^*}^{\pm}}$  
as 0.9089,
when the vacuum masses for the charged open 
charm vector and pseudoscalar mesons (${D^*}^{\pm}$ 
and $D^{\pm}$) are used in equation (\ref{decay_VP}). 
However, in the presence of a magnetic field, 
the masses of these charged mesons 
have contributions from the Landau levels (dominantly 
from the lowest Landau level in the presence of strong 
magnetic fields) and are given by 
$m_{D^\pm}(eB)=\sqrt {{m_{D^\pm}^{vac}}^2 +eB},\,
m_{{D^*}^\pm}(eB)=\sqrt {{m_{{D^*}^\pm}^{vac}}^2 -eB}$.
The coupling parameter $g_{PV}\equiv
g_{D^{\pm}{D^*}^{\pm}}$, as obtained from
the decay width of the process 
${D^*}^{\pm} \rightarrow D^{\pm} \gamma$, 
thus turns out to be dependent on the magnetic field.
The value of this parameter for $D^\pm-{D^*}^\pm$ mixings
is modified from the value of 0.9089 for zero 
magnetic field to around 1.484, 2.0587 and 5.7337
for the values of $eB/m_\pi^2$ of 4, 6 and 10 respectively.
The value of this coupling parameter for a given
magnetic field and the masses of $D^\pm$ and
${D^*}^\pm$ mesons including the lowest Landau 
level contributions, are used in equation (\ref{mpv_long}) 
to compute the masses $m^{(PV)}_{V^{||},P}$, due to 
the pseudoscalar-vector meson mixing,  
of the pseudoscalar meson $P\equiv D^{\pm}$ and 
longitudinal component of the vector meson 
$V \equiv {D^*}^{\pm}$,
in the presence of the magnetic field. 
The masses for these charged mesons 
obtained from the PV mixing
in the presence of a magnetic field,
have in addition the lowest Landau level contributions, 
as given by equation (\ref{m_LL}).  
The large increase in the value of the 
$D^\pm-{D^*}^\pm$ mixing parameter at high 
magnetic fields, as we shall see later, 
is observed to modify appreciably
the masses of the charged open charm mesons
(and hence the decay widths involving these charged mesons),
as compared to the case when the mixing parameter is taken 
to be fixed at the zero magnetic field value.

We next consider the mixing effect between the neutral $D$ 
and $D^*$ mesons in the presence of a magnetic field. 
The value of $g_{D^0 {D^*}^0}$ is needed to
study the mixing of the ${D^*}^0$ and ${D}^0$ mesons,
which can be obtained from the decay width 
of ${D^*}^0 \rightarrow D^0 \gamma$. This partial decay width 
is 35.3 $\%$ of the total width of ${D^*}^0$ \cite{pdg_2019_update}.
However, its value is not known, as the total width of ${D^*}^0$ 
is still not measured experimentally with accuracy 
\cite{pdg_2019_update}. 
The decay of ${D^*}^0$ comprises of the
decay modes ${D^*}^0 \rightarrow D^0 \pi^0$ and 
${D^*}^0 \rightarrow D^0 \gamma$ with the branching
ratio of 64.7:35.3 \cite{pdg_2019_update}.
In the present work, 
the value of the radiative decay width of ${D^*}^0$
is estimated in a similar manner as was done
in Ref. \cite{Gubler_D_mag_QSR}.
The coupling constant for ${D^*}^0 \rightarrow D^0 \pi^0$
is assumed to be the same
as that of the decay ${D^*}^+ \rightarrow D^+ \pi^0$,
the latter as calculated from its observed decay width.
This value of the coupling constant is used to calculate
the decay width of ${D^*}^0 \rightarrow D^0 \pi^0$,
and the decay width of ${D^*}^0 \rightarrow D^0 \gamma$,
is then obtained from the observed branching ratio of these
two decay modes. It might be noted here that in 
Ref. \cite{Gubler_D_mag_QSR}, the decay width of the
process $D^*\rightarrow D\pi$ was calculated using a
phenomenological Lagrangian interaction
${\cal L}_{int}\sim g \pi (\partial ^\mu D){D^*}_\mu$
and the value of the coupling constant, $g$,
was assumed to be same for the processes, 
${D^*}^0 \rightarrow D^0 \pi^0$ and
${D^*}^+ \rightarrow D^+ \pi^0$, 
using isospin symmetry. In the present work, we use
the values of the coupling strength $\gamma_{D^*}$
as given in the expression for the decay width
of ${D^*} \rightarrow D\pi$ (given by equation
(\ref{gammadstr})) calculated within the field theoretic
model of composite hadrons, to be same for these two processes.
The observed decay width of $\Gamma ({D^*}^+\rightarrow D^+ \pi^0)$
as 25.6 KeV \cite{pdg_2019_update}
yields the value of $\gamma_{D^*}$ to be 4.27. 
Using this value for $\gamma_{D^*}$, the decay
width of $\Gamma ({D^*}^0\rightarrow D^0 \pi^0)$
is obtained as 35.9 keV. This gives the 
value of the decay width $\Gamma ({D^*}^0\rightarrow D^0 \gamma)$
as 19.593 KeV, using the measured branching ratio 
of $\Gamma ({D^*}^0\rightarrow D^0 \pi^0):
\Gamma ({D^*}^0\rightarrow D^0 \gamma)$ to be 64.7:35.3
\cite{pdg_2019_update}.
Using equation (\ref{decay_VP}), the value 
of the coupling parameter, $g_{VP}=g_{{D^*}^0D^0}$ 
is obtained as 3.426. This value of the coupling parameter
may be compared to the value of 3.6736 
Ref. \cite{Gubler_D_mag_QSR}, evaluated using
the phenomenological Lagrangian interaction 
$\sim g \pi (\partial ^\mu D){D^*}_\mu$
and using the  branching ratio of $\Gamma ({D^*}^0\rightarrow D^0 \pi^0):
\Gamma ({D^*}^0\rightarrow D^0 \gamma)$ as 61.9:38.1.
The value of the mixing strength parameter, 
$g_{{D^*}^0 D^0}=3.426$ as estimated in the present work,
is used to study the effects of the mixing on the masses
of the $D^0$ and ${D^*}^0$ mesons
in the presence of a magnetic field.
The effect of the magnetic field on the decay
${D^*}^+\rightarrow D^0 \pi^+$ is also studied in
the present work. The value for $\gamma_{D^*}$ for this channel 
is obtained to be 5.94, from the measured value of the vacuum 
decay width of ${D^*}^+\rightarrow D^0 \pi^+$ as 56.46 keV 
\cite{pdg_2019_update}.

The masses for the charged and neutral open charm mesons 
are plotted as functions of $eB/m_\pi^2$ in figure \ref{md_PV}.
As has already been mentioned, 
the parameter for $D^+-{D^*}^+$ mixing is calculated 
using equation (\ref{decay_VP}),
from the observed value of the radiative decay width 
of ${D^*}^+\rightarrow D^+ \gamma$ and, this parameter
is dependent on the magnetic field, due to the Landau
level contributions to the masses for these charged mesons 
in the presence of a magnetic field. 
The  effects of the pseudoscalar-vector meson mixing
(marked as PV) on the masses of these mesons are shown
for the charged and neutral mesons in panels 
(a) and (b) respectively, when the $D^+-{D^*}^+$ mixing
parameter is fixed at the zero magnetic field value.
For this case, the mixing is observed to lead to a monotonic
decrease (increase) in the mass of the $D$
(longitudinal component of ${D^*}$) meson.
For the neutral $D$ and ${D^*}$ mesons, the shifts in the masses 
are observed to be much larger due to the mixing effects,
as compared to the case of the charged mesons. This is due to the
much larger value of the coupling parameter, $g_{VP}$
of 3.426 for $D^0 {D^*}^0$ mixing as compared to the
value of 0.9089 (for zero magnetic field case) 
for the $D^+-{D^*}^+$ mixing.
In the present work, the masses of the charged $D$ and $D^*$ mesons 
have contributions from the lowest Landau level, 
as given by equation (\ref{m_LL}). 
These contributions, shown as the dot-dashed 
lines, lead to a rise (drop) in the mass of the $D^+$ 
(${{D^*}^+}$) meson. 
In panel (a), the masses due to the combined effects 
of including the Landau level as well as the mixing 
contributions are shown for the $D^+$ and ${D^*}^+$ respectively. 
These masses are observed to lead to an initial increase (drop)
for the $D^+$ (${D^*}^+\,||$) upto around $eB=5 m_\pi^2$,
followed by a slow decrease (increase) as the magnetic field
is further increased, when the mixing effects dominate over the
Landau level contributions. 
When the mixing parameter for the charged open charm mesons
is fixed at the zero magnetic field value of 0.9089,
the mass of the ${D^*}^+\,^{||}$ ($D^+$) (in MeV) is modified
from its vacuum value 2010.26 (1869.65) to
2014 (1866) and 2024 (1856.8) 
at $eB=5m_\pi^2 $ and $eB=10 m_\pi^2$
respectively in the presence of only mixing effect
and to 1991.4 (1890.3) and 1991.2 (1892.1), 
when the Landau level contributions are also taken into account.
The masses of the charged open charm mesons are thus 
observed to vary only marginally for $eB$ larger 
than 5 $m_\pi^2$.
On the other hand, the mass of  neutral 
${D^*}^0\,^{||}$ (${D}^0$)
(in MeV) is observed to be modified from the vacuum value 
of  2006.9 (1864.83) to  2049.4 (1826.1) and 2129.9 (1757)
at $eB=5m_\pi^2 (\sim 0.1 {\rm GeV}^{-2})$ 
and $eB=10 m_\pi^2 (\sim 0.2 {\rm GeV}^{-2})$
respectively in the presence of the mixing effect.
The shifts in the masses of the neutral open charm
mesons are thus observed to be much larger than for
the charged ${D^*}^+$ and ${D}^+$ mesons,
as is observed in figure \ref{md_PV}, when the 
charged $D^*-D$ mixing parameter is fixed at its value at $eB=0$. 
However, there are observed to be dominant modifications
when the magnetic field dependence of this paramter 
is considered. Above a value of the magnetic field
of around 4 $m_\pi^2/e$, there is observed to large 
increase (drop) in the mass of the ${D^*}^+ (D^+)$ meson.
The values of the mass of the ${D^*}^+(D^+)$, accounting for both
the PV mixing as well as Landau quantization effects, 
are observed to be modified from 1989.5 (1892.5) and 1991.2 (1892.1) 
to 2015.7 (1867.9) and 2250.7 (1673.9)
for $eB=6 m_\pi^2$ and  $eB=10 m_\pi^2$,
when the magnetic field dependence
of the ${D^*}^+-D^+$ mixing parameter is taken into account. 
These are observed to modify the decay widths
of the charged ${D^*}$ meson as well as $\Psi(3770)
\rightarrow D^+D^-$, as we shall later.

The decay widths of $D^*\rightarrow D\pi$ are computed 
using a field theoretic model of composite hadrons.
The Landau level contributions
modify the masses of the charged $D$ and $D^*$ mesons,
whereas, the pseudoscalar vector meson mixing modifies the
masses of the longitudinal component of the $D^*$ meson
as well as of the $D$ mesons.
The decay width of $D^*\rightarrow D\pi$ is given 
by equation (\ref{gammadstr_mix}), which has 
contributions from mixing effects from the longitudinal
$D^*$ meson.  The dependence of the
decay width given by equation (\ref{gammadstr_mix}) 
on the masses of the $D$, $D^*$
and $\pi$ are through the center of mass
momentum $|{\bf p}|$ 
given in terms of the masses 
of the $D$ and $D^*$ mesons, including 
the Landau level contributions for the charged mesons,
whereas, the second term has the masses of these
mesons including the mixing effects as well.
The center of mass momentum,  $|{\bf p}|$ is given 
by equation (\ref{modp}),
in terms of the masses of the decaying, $D^*$ meson
and the produced mesons, $D$ and $\pi$.

The effects of the magnetic field on the decay widths
of the charged ${D^*}^+$ and ${D^*}^0$ are shown in
panels (a) and (b) in figure \ref{dwFT_Dstr}.
When the $PV$ mixing parameter is fixed at
its zero magnetic field value for the charged open charm mesons
($g_{PV}=g_{PV}(eB=0)$),
in the presence of only the pseudoscalar-vector meson
mixing (marked as PV), the decay width of 
${D^*}^+ \rightarrow D^+ \pi^0$ is observed to 
be modified only marginally as compared to the
case when these effects are not taken into account
(the decay widths remain at their vacuum values
as shown by the dotted line). This is because 
the small value of the coupling parameter,
$g_{PV}(eB=0)$=0.9089 for the $D^+-{D^*}^+$ mixing,  
leads to moderate modifications in the masses 
of the ${D^*}^+$ and $D^+$ mesons.
However, the decay width of ${D^*}^+ \rightarrow D^0 \pi^+$ 
is observed to increase appreciably with the magnetic field,
when we consider only the mixing effect,
from its vacuum value of 56.46 keV to
a value of 0.6569 (3.991) MeV at the value of eB
as $5m_\pi^2$ (10 $m_\pi^2$).
In the presence of $PV$ mixing effects as well as 
Landau level contributions,
when the magnetic field dependence of the ${D^*}^+-D^+$
mixing parameter is taken into account, there is observed to
be steeper rise of the decay width ${D^*}^+\rightarrow D^0 \pi^+$,
as compared to the the case when this parameter is fixed
at the zero magnetic field value, as can be seen from
figure {\ref{dwFT_Dstr}. However, the decay width
of ${D^*}^+\rightarrow D^+\pi^0$ remains almost unchanged,
for the cases of $g_{PV}(eB)$ and $g_{PV}(eB=0)$,
due to the marginal magnetic field dependence 
of this parameter for upto $ eB \sim 0.5 m_\pi^2$
(beyond which the decay width drops appreciably).
In the presence
of Landau effects, there is observed to be an initial 
drop in the width of ${D^*}^+ \rightarrow D^0 \pi^+$ 
followed by a rise as the magnetic field is further increased. 
In the panel (b), the logarithm of the decay width
of ${D^*}^0 \rightarrow D^0 \pi^0$ is plotted
as a function of $eB/m_\pi^2$. The decay width 
of ${D^*}^0\rightarrow D^0\pi^0$ is observed
to have large contribution from the mixing effects.
This is due to the large positive (negative) shifts
in the mass of the neutral vector ${D^*}^0$ (${D}^0$) 
in the presence of a strong magnetic field.
The decay widths for the modes ${D^*}^+\rightarrow D^0\pi^+$
as well as  ${D^*}^0\rightarrow D^0\pi^0$ are observed to 
have significant contributions for large values of the
magnetic fields. These should have implications in the
enhancement of the neutral $D$ mesons as compared
to the charged $D^+$ meson. 

In figure \ref{dwFT_3770_vac_ddst_spinmix}, the dependence
of the decay widths of $\Psi(3770)\rightarrow D\bar D$
on the magnetic field are shown, for the cases when 
the contributions from PV mixing to the masses of 
the charmonium as well as $D$ and $\bar D$ mesons 
are considered. These are compared with the case
(shown as dotted lines) when the mass modifications 
from the pseudoscalar-vector meson
mixing are not taken into account for the open charm mesons
in panel (b) of figure \ref{dwFT_3770_vac_ddst_spinmix}
\cite{charmonium_PV_amspm}.
In panel (a), the decay widths of $\Psi(3770)$ 
to the charged and neutral $D\bar D$ pair are shown
for the case when the PV mixing contributions are not taken
into account for the charmonium as well as open charm mesons,
but only the Landau level contributions are considered
for the charged $D^{\pm}$ mesons. Due to the positive
Landau level contributions to the masses of the charged
pseudoscalar mesons,  there is observed to be a decrease
of the decay width for the $D^+D^-$ final state,
which is observed to vanish for $eB \ge 3 m_\pi^2$. 
On the other hand, the final state $D^0 \bar {D^0}$ 
decay width remains unaffected by the magnetic
field (as there are no Landau level contributions
for the neutral open charm mesons). Fig 3(b) shows the
results for the decay widths of $\Psi(3770)$
accounting for the PV mixing contributions for the
masses of the charmonium state as well as open charm mesons,
in addition to the Landau 
level contributions for the charged open charm mesons.
It is observed that the PV mixing contributions 
to the masses of the final state open charm mesons, 
along with the PV mixing contributions to the
mass of $\Psi(3770)$ (due to mixing with $\eta_c'$)
lead to larger values of the decay widths.
As has already been mentioned, the Landau level contributions
are considered for the charged open charm 
mesons in the presence of the magnetic field. 
As may be seen from figure 1(a), 
while accounting for the PV mixing 
contributions along with the Landau level contributions,
there is observed to be an initial increase (drop) in the 
mass of the charged $D^+$ (${{D^*}^+}^{||}$) with
increase in the magnetic field, due to the Landau level
contribution dominate over the PV mixing contributions.
This leads to a decrease in the mass difference between the 
$D^+$ and ${{D^*}^+}^{||}$ with rise in the magnetic field,
due to which the PV mixing effects start becoming more important
since the mass splitting between the pseudoscalar and longitudinal
componenent of the vector mesons from PV mixing is 
inversely proportional to the mass difference of the mesons
\cite{Gubler_D_mag_QSR,charmonium_PV_amspm}.
As the magnetic field is further increased, the PV mixing
contributions dominate over the Landau level contributions.
At higher values of the magnetic field, the mass of the
charged pseudoscalar meson is observed to drop.
This is observed as enhancement of the decay width
of $\Psi(3770)\rightarrow D^+D^-$ when the PV mixing
effects are also taken into account for the open charm mesons,
as compared to when the PV mixing effects are considered
only for the charmonium state (due to $\Psi(3770)-\eta_c'$
mixing). The decay width of $\Psi(3770)\rightarrow D^0 \bar {D^0}$
is observed to have a positive contribution from the 
$D-D^*$ mixing, as may be seen from the panel (b) of figure
\cite{charmonium_PV_amspm}. There is observed to be a drop 
in the decay width with the neutral $D\bar D$ final state,
with further increase in the magnetic field. 
The decay width (in MeV) of $\Psi(3770)$ to the neutral $D\bar D$
is observed to be around 39.1 (33.2) for $eB$= 6 (10) $m_\pi^2$,
which is much larger as compared to the values of 6.4 (10.2)
for the $D^+D^-$ final state, when the ${D^*}^\pm -D^\pm$ 
mixing parameter is fixed at its zero magnetic field value,
and 10.8 (3.1) when the magnetic field dependence of the
charged $D^*-D(\bar {D^*}-{\bar D}$ mixing is taken into account. 
The much larger value of the decay width to the
neutral $D\bar D$ pair as compared to the $D^+D^-$ final state,
may be observed as the neutral $D\bar D$ mesons 
to be much more profusely produced 
from decay of charmonium states 
as compared to the charged $D\bar D$ mesons 
in the presence of strong magnetic fields.

\section{Summary}

The masses of the $D$ and $D^*$ mesons are studied 
in the presence of magnetic fields, 
taking the effects of the mixing of the pseudoscalar
and vector mesons into consideration. 
The charged open charm mesons have additionally 
the Landau level contributions. The masses of the $D$
(and $\bar D$) and the vector $D^*$ mesons 
in the presence of a magnetic field
are used to study the decays of the
charged and neutral $D^*$ mesons to $D\pi$,
as well as to study the decay width of $\Psi(3770)\rightarrow
D\bar D$.
 For the charmonium decay width, the mass
modification of $\Psi(3770)$ in the presence
of a magnetic field is due to its mixing with the
pseudoscalar mesons, $\eta_c'$. The decay widths are computed
by using a composite model for the hadrons with 
quark and antiquark constituents, 
using the light quark-antiquark pair creation term 
of the free Dirac Hamiltonian for the constituent quarks.
The matrix element of this term between the initial
and final states is used to calculate the decay width.
The matrix element is multiplied with a strength parameter 
for the quark pair creation, which is fitted from the observed vacuum
decay widths for the specific decay process ($D^* \rightarrow D\pi$
and $\Psi(3770)\rightarrow D\bar D$). 
For the charged $D^*-D$ mixing, the parameter $g_{PV}$
depends on the magnetic field, because of the Landau
contributions to the masses of these charged mesons
in the presence of a magnetic field. The ${D^*}^\pm-D^\pm$
mixing parameter is observed to increase appreciably 
with increase in the magnetic field. This leads to 
dominant modifications of the charged open charm mesons,
and hence on the decay widths of the charged $D^*\rightarrow D\pi$,
as well as of $\Psi(3770)\rightarrow D^+D^-$ at large
values of the magnetic field.
The $\Psi(3770)$
decays dominantly to the neutral open charm mesons
as compared to $D^+D^-$ mesons,
in the presence of strong magnetic fields.
The created magnetic fields in the peripheral ultra-relavtistic 
heavy ion collsion experiments, e.g., at RHIC and LHC, 
are huge, and the matter resulting from the high energy collision
is of (extremely) low density. For zero (extremely small) density,
the lowest charmonium state which can decay to $D\bar D$ 
is $\Psi(3770)$. The present work of the
study of the effects of magnetic field on the masses
of the open charm mesons, $D$, $\bar D$, $D^*$ and
$\psi(3770)$, and their subsequent effect on the 
decay widths of $D^*\rightarrow D\pi$ and $\Psi(3770)\rightarrow
D\bar D$ can be of relevance to the observables, e.g., the
production of the open charm mesons and charmonium
states in peripheral ultra-relativistic heavy ion collision 
experiments.

One of the authors (AM) is grateful to ITP, University of Frankfurt,
for warm hospitality and 
acknowledges financial support from Alexander von Humboldt Stiftung 
when this work was initiated. 


\end{document}